\documentclass{ptapap}

\usepackage{footnote}
\usepackage{url}

\author{Arkadiusz Hypki}[UL,CAMK]
\author{Anthony Brown}[UL]
\affil[UL]{Leiden Observatory, Leiden University, PO Box 9513, NL-2300 RA
Leiden, the Netherlands, ahypki@strw.leidenuniv.nl} \affil[CAMK]{Nicolaus
Copernicus Astronomical Center,   Bartycka 18, 00--716 Warsaw, Poland}

\title{Gaia archive}

\begin{document}

\maketitle

\begin{abstract}

The Gaia archive is being designed and implemented by the DPAC Consortium.
The purpose of the archive is to maximize the scientific exploitation of the
Gaia data by the astronomical community. Thus, it is crucial to gather and
discuss with the community the features of the Gaia archive as much as possible.
It is especially important from the point of view of the GENIUS project to
gather the feedback and potential use cases for the archive. This paper
presents very briefly the general ideas behind the Gaia archive and presents
which tools are already provided to the community.

\end{abstract}

\section{Introduction}

Gaia is an ambitious mission of ESA. Its goal is to chart a
three-dimensional map of the Galaxy \citep{Gaia2008IAUS..248..217L}. It will
provide extraordinarily precise measurements for over one
billion stars in the Milky Way including astrometry:
stellar positions, parallaxes, and proper motions; photometry:
photometric magnitudes in different spectral bands and at each epoch;
spectroscopy: radial velocities and astrophysical parameters. Additionally, it
will observe hundreds of thousands of solar system bodies, extragalactic
sources, distant quasars and much more. Gaia will provide unprecedented
positional and radial velocity measurements with the accuracies needed to
produce a stereoscopic and kinematic census of about one billion stars in our
Galaxy and throughout the Local Group. This is by far the most advanced
astrometric mission ever launched.

The GENIUS project is an international effort to support ESA to design and
develop the Gaia archive taking into account the needs of the user community and to
maximize the scientific return from the Gaia mission. The Gaia data processing,
of which final result will be the Gaia archive, is the responsibility of the
DPAC consortium --- a multinational European group of scientists and engineers.
The design and implementation of this archive was opened by ESA to
participation by the European community. GENIUS aims to significantly
contribute to the development of the Gaia archive. It takes into account
the input from the astronomical community, the interoperability with already
existing and coming astronomical missions, and the cooperation with nanoJASMINE
and JASMINE -- the only other astrometric missions under development.

The typical use cases of astronomical problems which researchers would like to
solve using Gaia archive were already gathered -- long before the Gaia launch.
However, feedback from the researchers and discussion about already
implemented tools for the Gaia archive is still very important. It helps to choose
next goals and tweak the current software developments plans in order to make
the Gaia archive even more useful and more accessible for a broader range of
users.

\section{Tools for the Gaia archive}

There is already available a number of tools which are connected to the Gaia
archive, fully working and ready to serve real scientific data (currently Gaia
archive contains only measurements generated from a numerical model of the Milky
Way). 

The main interface to access the Gaia archive is GACS -- Gaia Archive Core
Systems\footnote{\url{http://gaia.esac.esa.int/archive/}}. It is a web portal
which allows users to specify a query, run it within the European Space
Astronomy Centre (ESAC) infrastructure and after completion view the results.
The query language is called ADQL, which stands for Astronomical Dataset Query Language \citep{ADQL2007ASPC..382..381P}. It is
an SQL-like language designed to search and join astronomical data sets, adopted
as a standard and widely used across the whole Virtual Observatory
initiative\footnote{\url{http://www.ivoa.net/}} -- an international effort to
make various astronomical data sets and other resources accessible in a
standardized way. The Gaia archive implements all needed protocols (TAP, SAMP)
which makes it fully accessible from any VO compatible tool.

Other interesting tools which are already connected to Gaia archive and allow to
exploit the Gaia data are
Topcat\footnote{\url{http://www.star.bris.ac.uk/~mbt/topcat/}}
\citep{Topcat2005ASPC..347...29T} and
Vaex\footnote{\url{http://www.astro.rug.nl/~breddels/vaex/}}. Topcat is a
graphical tool which allows to view and edit a tabular data. It is especially
designed to help astronomers to analyse and manipulate the source catalogues.
It has implemented all the major formats and protocols widely used in astronomy,
like FITS files, VOTable and thus it is seamlessly connected to the Gaia
archive. Vaex is a graphical tool to visualize and explore large tabular
datasets. It handles tables with the number of rows of the order $10^9$ in
seconds, thus it is suitable to explore the Gaia archive. Moreover, Vaex, also
implements the VOTable protocol which makes it integrated with the Gaia archive
out-of-the-box.

Within the GENIUS project there are also designed and implemented tools able to
handle really huge datasets ($\sim$~PBs of data) in order to compare them
against the whole or a large fraction of the Gaia archive. Simple bash/python
scripts for data analysis of such huge datasets are not a reasonable solution
here.
There is a need to have a more advanced tool. One of such tools is the software
called BEANS\footnote{\url{http://beanscode.net}}.
It allows to query, aggregate and visualize huge data sets much easier and
faster. It uses Apache Cassandra\footnote{\url{http://cassandra.apache.org}} as
a database (one of the best solutions from emerging NoSQL technology movement),
PigLatin\footnote{\url{http://pig.apache.org}} as a scripting language to deal
with distributed data analysis, and the library called
D3\footnote{\url{http://d3js.org}} which is an example on how computer sciences
and art sciences can work together creating non-standard plots which at once are
powerful, interactive, clean and easy to read.
The BEANS software is one of the first attempts to adopt NoSQL data analysis
techniques to astrophysics. It is written in a very general form, so it can be
used in any field of research (economy, physics, biology etc.), commercial
companies or other open source projects. All the necessary code which connects
BEANS with the GACS portal is already implemented as a plugin. It allows to
write an ADQL query which later one one can compare within BEANS with e.g.
numerical simulations of Milky Way or globular star clusters. The first public
version of the BEANS software, together with GACS plugin, is planned for release
in early 2016.
Then, the software will be constantly improved taking into account feedback from
users.

The discussion on already existing features of the implemented software and
presenting a new use case scenarios for the Gaia archive is still possible.
The list of the current use cases gathered in the community one can find in the document ``Gaia
data access scenarios
summary''\footnote{\url{http://www.rssd.esa.int/doc_fetch.php?id=3125400}}
(GAIA-C9-TN-LEI-AB-026). On the Gaia Research for European Astronomy Training (GREAT) wiki
pages\footnote{\url{http://great.ast.cam.ac.uk/Greatwiki/GaiaDataAccess/GdaScenariosFeedback}}
one can find the information on how to suggest more use cases.

\section{Summary}

The first public data release of Gaia data is planned for Summer 2016. In order
to maximize the scientific output of the Gaia mission it is crucial to put a lot
of effort to create and implement a set of tools ready to allow scientists to
analyze the Gaia data. That is the main goal of the GENIUS project -- take the
input from the community and help to implement software which would support
scientific research as much as possible. The Gaia archive is accessible through
the Virtual Observatory, which in the recent years became a standard way to
distribute astronomical catalogues. There is already a number of tools
connected to the Gaia archive (Topcat, Vaex) and there are even more tools to come
(BEANS, GAVIP\footnote{\url{http://docs.gavip.science/#section2}}). The first
Gaia data are just around the corner and it is crucial to provide to the
community along the data the software which would allow to analyse the data
easier.

\acknowledgements{The research leading to these results has received funding
from the European Community's Seventh Framework Programme (FP7-SPACE-2013-1)
under grant agreement n.~606740.}

\bibliographystyle{ptapap}
\bibliography{ptapapdoc}

\begin{thebibliography}{3}
\providecommand{\natexlab}[1]{#1}
\providecommand{\url}[1]{\texttt{#1}}
\providecommand{\urlprefix}{URL }
\providecommand{\eprint}[2][]{\url{#2}}

\bibitem[{{Lindegren} et~al.(2008)}]{Gaia2008IAUS..248..217L}
{Lindegren}, L., et~al., \emph{{The Gaia mission: science, organization and
  present status}}, in W.~J. {Jin}, I.~{Platais}, M.~A.~C. {Perryman} (eds.)
  IAU Symposium, \emph{IAU Symposium}, volume 248, 217--223 (2008)

\bibitem[{{Plante}(2007)}]{ADQL2007ASPC..382..381P}
{Plante}, R., \emph{{Chapter 36: The Astronomical Dataset Query Language
  (ADQL)}}, in M.~J. {Graham}, M.~J. {Fitzpatrick}, T.~A. {McGlynn} (eds.)
  Astronomical Society of the Pacific Conference Series, \emph{Astronomical
  Society of the Pacific Conference Series}, volume 382, 381 (2007)

\bibitem[{{Taylor}(2005)}]{Topcat2005ASPC..347...29T}
{Taylor}, M.~B., \emph{{TOPCAT \& STIL: Starlink Table/VOTable Processing
  Software}}, in P.~{Shopbell}, M.~{Britton}, R.~{Ebert} (eds.) Astronomical
  Data Analysis Software and Systems XIV, \emph{Astronomical Society of the
  Pacific Conference Series}, volume 347, 29 (2005)

\end{thebibliography}

\end{document}